\documentclass[]{spie}  
\usepackage[latin1]{inputenc}
\usepackage[british]{babel}
\usepackage[]{graphicx}
\usepackage{amsmath,amssymb}
\usepackage{textcmds}
\usepackage{astro_bib_macro} 
\usepackage{multirow}
\usepackage{color}

\title{High-contrast imager for Complex Aperture Telescopes\\ (HiCAT):\\1. Testbed design} 

\author{Mamadou N'Diaye\supit{a}*, Elodie Choquet\supit{a}, Laurent Pueyo\supit{a,b}, Erin Elliot\supit{a}, Marshall D. Perrin\supit{a},\\J. Kent Wallace\supit{c}, Tyler Groff\supit{d}, Alexis Carlotti\supit{e}, Dimitri Mawet\supit{f,c}, Matt Sheckells\supit{b},\\Stuart Shaklan\supit{c}, Bruce Macintosh\supit{g}, N. Jeremy Kasdin\supit{d} and R\'emi Soummer\supit{a\dag}
\skiplinehalf
\supit{a} Space Telescope Science Institute, 3700 San Martin Drive, Baltimore, MD 21218, USA\\
\supit{b} Dept. of Physics and Astronomy, Johns Hopkins University, Baltimore, MD 21218, USA\\
\supit{c} Jet Propulsion Laboratory, California Institute of Technology, Pasadena, CA 91109, USA\\
\supit{d} Mechanical and Aerospace Engineering, Princeton University, Princeton, NJ 08544, USA\\
\supit{e} Institut de Plan\'etologie et d'Astrophysique de Grenoble (CNRS/UMR 5274), F-38041 Grenoble Cedex 9, France\\ 
\supit{f} European Southern Observatory, Vitacura, 763 0355, Santiago, Chile\\
\supit{g} Lawrence Livermore National Laboratory, 7000 East Ave L-210, Livermore, CA 94040, USA\\
}

\authorinfo{*E-mail: mamadou@stsci.edu, \dag: Principal Investigator}

\begin{document} 
  \maketitle 

\begin{abstract}
Searching for nearby habitable worlds with direct imaging and spectroscopy will require a telescope large enough to provide angular resolution and sensitivity to planets around a significant sample of stars. Segmented telescopes are a compelling option to obtain such large apertures. However, these telescope designs have a complex geometry (central obstruction, support structures, segmentation) that makes high-contrast imaging more challenging. We are developing a new high-contrast imaging testbed at STScI to provide an integrated solution for wavefront control and starlight suppression on complex aperture geometries. We present our approach for the testbed optical design, which defines the surface  requirements for each mirror to  minimize the amplitude-induced errors from the propagation of out-of-pupil surfaces. Our approach guarantees that the testbed will not be limited by these Fresnel propagation effects, but only by the aperture geometry. This approach involves iterations between classical ray-tracing optical design optimization, and end-to-end Fresnel propagation with wavefront control (e.g. Electric Field Conjugation / Stroke Minimization). The construction of the testbed is planned to start in late Fall 2013.
\end{abstract}



\section{INTRODUCTION}\label{sec:intro}
Direct imaging and spectroscopy of exoplanets represent an exciting opportunity for comparative exoplanetology to understand the formation and evolution of circumstellar environments. New instruments with high-contrast imaging capabilities will soon be in operation on the ground with 8\,m-class telescopes (Gemini/GPI\cite{2008SPIE.7015E..31M}, VLT/SPHERE\cite{2008SPIE.7014E..41B}, Subaru/ScExAO\cite{2010SPIE.7736E..71G}) and in orbit with the 6.5\,m James Webb Space Telescope\cite{2010ASPC..430..167C}, allowing for the detection and spectral analysis of hot, gaseous planets around nearby stars.

The search for fainter companions such as habitable worlds by direct imaging will require a telescope large enough to provide angular resolution and sensitivity to planets around a significant number of nearby stars. Segmented telescopes constitute a compelling option to address these questions. However, their aperture geometry (segmentation, central obstruction and spider vanes) makes high-contrast imaging more challenging. We are developing a testbed at the Space Telescope Science Institute (STScI) to provide an integrated solution for wavefront sensing and control and starlight suppression strategies on complex aperture geometries. This testbed is located at the Russell B. Makidon Optics Laboratory, a new state-of-the art facility at STScI dedicated to developing technologies for future space missions. 

Our initial contrast goal for the testbed in air is $10^{-7}$ in the visible. The main focus is developing methods for segmented aperture control that can be later implemented and tested at higher contrast e.g. at HCIT\cite{2004SPIE.5487.1330T}. To study the effects of aperture geometries and understand the ultimate contrast capabilities, we have designed a testbed with a reduced impact of bench optics, in particular the amplitude-induced errors from the propagation of out-of-pupil surfaces. Our methodology is based on an hybrid approach that combines optical ray tracing and end-to-end Fresnel simulations. In this paper, we present this approach, the results for the optics specifications and the final design of the testbed with all its features.

\section{APPROACH FOR DESIGN DEVELOPMENT}\label{sec:optimization}
This section presents our methodology for the development of the testbed optical design. We first detail the different constraints and describe our iterative approach for the design.

\subsection{Optical design constraints}
\begin{figure}[!ht]
\centering
\includegraphics[height=5cm]{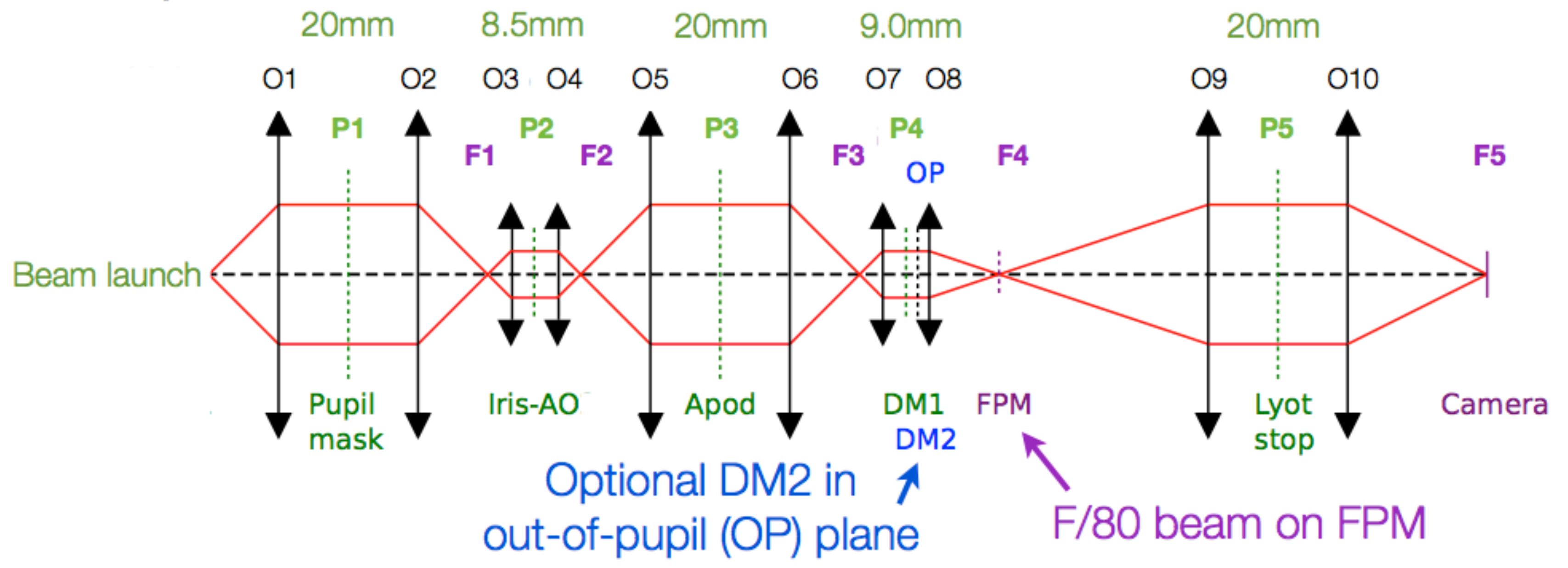}
\caption{Schematic linear representation of the high-contrast imaging testbed (light propagation goes from the left to the right). This layout involves 5 pupil planes (P), 5 focal planes (F) and 1 out-of-pupil (OP) plane for an optional deformable mirror, leading to ten optics (O) for a design with collimated beams. The beam size in each pupil plane is indicated on top. The apodizer (Apod), focal plane mask (FPM), and Lyot stop constitutes the components of the coronagraphic system. The terms Iris-AO and DM refer to the segmented and the face-sheet deformable mirrors. The actual testbed implementation is fully reflective with mirrors.}
\label{fig:testbed_conceptual_design}
\end{figure}

The testbed will be installed on a 2.4\,m$\times$1.2\,m table and equipped with laser sources to work in the visible. Several components have been identified for the testbed optical layout to carry out our investigations. We list here the main ones that put constraints on the design: 
\begin{itemize}
\item An entrance pupil mask with central obstruction and spiders to define an aperture shape. We set the size of the pupil mask to 20\,mm to ease the manufacture of mask struts (of the order of 1\% of the diameter), while enabling working with standard one inch optics.
\item A 37-segment Iris-AO\footnote{http://www.irisao.com} MEMs deformable mirror with 1.4 mm (vertex to vertex) hexagonal segments controllable in tip, tilt, and piston, conjugated to the entrance pupil to provide a pupil with complex geometry. We set a 8.51\,mm beam size on the deformable mirror to capture the circumscribed circle of all the segments. The gaps between segments range between 10 and 12\,$\mu$m. 
\item A diffraction suppression system based on the architecture of the apodized pupil Lyot coronagraph (APLC)\cite{2002A&A...389..334A,2005ApJ...618L.161S,2009ApJ...695..695S}. The design includes an apodizer, a reflective focal plane mask (FPM) and a Lyot stop. The apodizer and Lyot stop have the same size as the pupil mask (20\,mm) for symmetry reasons and manufacturing aspects. We set a F/80 beam focal ratio on the FPM to constrain the distances between the subsequent optics within the table dimensions and allow for a mask with reasonable physical size to ease manufacture, with a diameter in the range 300 to 500 microns. 
\item A Boston Micromachines\footnote{http://www.bostonmicromachines.com} deformable mirror (kilo-DM) for both wavefront control (correction of the residual aberrations) and wavefront shaping (generation of a dark hole inside the PSF). This DM shows 33 actuators across the 9.9\,mm lateral dimension. We set a 9.0\,mm beam size on the DM, capturing 30 actuators across the beam to work with a large number of actuators while keeping flexibility on the DM area selection with the beam.
\item A flat mirror immediately after the kilo DM for an easy upgrade towards the implementation of a future second Boston kilo-DM. With a second DM, we will be able to apply active correction of the aperture discontinuities (ACAD)\cite{2013ApJ...769..102P} and realize a two-DM wavefront shaping for amplitude and phase control. The distance between the two DMs has been set to 300\,mm.
\item A focal plane camera (CamF) with a motorized translation stage along the optical axis to capture coronagraphic images and perform classical or coronagraphic phase diversity methods\cite{2003JOSAA..20.1490D,2012OptL...37.4808S}.
\end{itemize}
Additional features have been identified outside the main optical path such as a camera for pupil imaging after the Lyot stop (CamP) or room for the implementation of low- and high-order wavefront sensing concepts, such as Zernike wavefront sensors \cite{2011SPIE.8126E..11W,2012SPIE.8450E..0NN,2013A&A...555A..94N}. All these devices and their parameters put several constraints on the design of the testbed. As a starting point, we focus on the main optical path of the design. A conceptual layout of the high-contrast imaging testbed is given in Figure \ref{fig:testbed_conceptual_design}, including 5 pupil planes and 10 powered optics. 

\subsection{Philosophy and contrast requirements}
The generation of a complex geometry pupil is ensured by the use of an Iris-AO 37-segmented deformable mirror and the presence of a centrally obstructed pupil mask with spiders. The discontinuities in the pupil (segment gaps, spiders, etc.) will be addressed as amplitude errors and corrected using wavefront control and wavefront shaping. Our goal is to reach an experimental $10^{-7}$ contrast at 3\,$\lambda/D$ (where $\lambda$ and $D$ denote the wavelength and aperture diameter) using initially a single continuous face-sheet DM (Boston kilo-DM). This initial requirement will ensure sufficient stroke availability to correct the aperture geometric features in large spectral broadband with two DMs in the future.

For the purpose of our study the contrast shall only be limited by the effects of the aperture geometry (segment gaps, phasing residual errors, and coronagraph design and optimization for that geometry) and shall not be limited by amplitude-induced errors from the propagation of out-of-pupil surfaces, the so-called Talbot effects. To achieve this goal an unconventional optimization of the optical design is required and is based on a minimization of the amplitude induced-errors from the Talbot effects. We build our approach on recent studies and analysis \cite{2006ApOpt..45.5143S,2007ApJ...666..609P,2010SPIE.7736E.109A}.
In order to prevent any limitation from Talbot effects, we place a requirement on the contrast contribution from Talbot-type amplitude errors to be one order of magnitude fainter, i.e. $10^{-8}$ inside a half-field dark zone defined from 3 to 10\,$\lambda/D$ with a 2\% bandpass. We assume monolithic pupils (no pupil struts nor segment phasing effects) and a perfect coronagraph to estimate the Talbot effects from the optics only.

The total phase to amplitude conversion occurs for a given optic located at $z=z_T/4$ from the pupil plane, with $z_T$ the Talbot distance given by:
\begin{equation}
z_T = 2 (D /f_{max})^2 /\lambda\,,
\end{equation}
where $f_{max}$ denotes the maximum spatial frequency in the pupil we want to control, here 10 cycles per pupil. Each optics of our design shall be at a distance $z$ from the pupil plane such that $z \ll z_T/4$, following the approach of Antichi et al. \cite{2010lyot.confE..61A,2010SPIE.7736E.109A}. This condition proves challenging for the optics close to MEMS DMs which have a small size ($\sim$\,9\,mm), associated with a short Talbot distance of 2.56\,m. The dimensions of the testbed table set limits to the maximum size and pupil magnification factors in our design. We adopt reflective optics for the design to ease broadband operations. This introduces constraints on how close mirrors can be placed to the DMs, so that the $z \ll z_T/4$ requirement is challenging.

\subsection{Iterative optimization of the design}
We adopt an hybrid approach combining analytical model, optics ray tracing and end-to-end Fresnel simulations to define the testbed design and determine the mirror specifications. Our first step deals with an optimization of the geometric optics design, including iterations of:
\begin{itemize}
\item optical ray tracing with Zemax. We start from a base focal length $f_0$ for the first optics of the system $O_1$ and from this value, we determine the focal lengths for the other optics, related to the magnification factors between two successive pupils. These parameters are introduced under Zemax to trace the ray optics design. 
\item mechanical constraint considerations (table size, basic opto-mechanics). We adjust the distances and the angles of the incident beams on reflective surfaces under Zemax to fit the testbed design into the table space, to ensure sufficient clearance for each component, and to avoid vignetting of the light beam.
\item analytical optimization of Talbot distances.  For each optics, the distance $z$ to the pupil plane is estimated and compared to the Talbot distance $z_T$ to check the requirement $z \ll z_T/4$ and minimize the Talbot effects.
\end{itemize}
We then proceed to the following analyses to define the mirror specifications:
\begin{itemize}
\item end-to-end testbed simulations with Fresnel propagation and starlight suppression algorithm (stroke minimization\cite{Pueyo:09}) to evaluate the contribution of amplitude-induced errors alone. 
\item maximization of the allowable wavefront errors (WFE) on each mirror while maintaining the contrast requirement of $10^{-8}$ in dark zone, for 2\% bandpass. 
\item maximization of the allowable low-order WFE (with spatial frequency content below 3 cycles/aperture) to include contribution for alignment error budget.
\end{itemize}
These steps are performed within two rounds of optimizations. In the first round, the goal is to determine preliminary surface error specifications for each mirror that reach the $10^{-8}$ contrast requirement. For each optics, the surface errors assume a conservative $f^{-2}$ power law power spectral density (PSD); Talbot coupling is negligible for low spatial frequencies. We consider a clear circular aperture and a perfect coronagraph to focus on the effects of the optics only. We perform a Fresnel propagation of the electric field from pupil plane to Lyot plane. Wavefront shaping is then applied on the resulting field to generate a PSF with half-field dark hole using the stroke minimization algorithm \cite{Pueyo:09}. We balance the level of RMS WFE on each mirror to obtain the specifications for each mirror that meet the final contrast requirement.
In the second round of optimizations, the goal is to increase the low-spatial frequency errors to account for the alignment error budget and relax the tolerances on the optics polishing. This is motivated also by the fact that the Talbot distances associated with low-order aberrations are very large, and therefore these low-spatial frequencies do not pose any issues for amplitude-induced errors and can be fully compensated by phase on the DM.  We repeat the same steps as in the first round of simulations, except that we simulate a partial pre-compensation of the phase for low-order aberrations. In practice this process corresponds to measuring the end-to-end wavefront error and correcting it  using the Boston DM. We then adjust the low-order wavefront error to refine the optics surface errors specifications.
We performed several iterations for both the geometric/Talbot distances and end-to-end optimizations to obtain an optical design with optics specifications that mitigate the Talbot effects under the opto-mechanical requirements.

\section{RESULTS}\label{sec:results}
\subsection{Minimization of the Talbot distances}
Table \ref{table:design_first_iteration} gives the initial and final sets of parameters for the testbed optics after ray-tracing design optimization (Zemax), and several iterations with end-to-end Fresnel simulations. Mirror sizes have been chosen to work with standard optical mounts of 25.4\, and 50.8\,mm (1" and 2") for simplicity and cost minimization. We define the field propagation distance $z$ between the optics and their closest pupil plane for all the mirrors except O8 for which the OP plane is considered. In the latter case, we assume a corrected beam after reflection on the deformable mirror in OP, leading us to consider Talbot effects for the residual errors from this plane.

The distance $z$ is expressed as a fraction of Talbot distance $z_T$ for both sets in the last two columns of Table \ref{table:design_first_iteration}. From the initial to the final set of parameters, we have reduced the distance of each optics to its reference plane from a quarter to smaller fractions of the Talbot length, allowing us to go from full to very partial phase-to-amplitude error conversion for each optics. With the final set, the optics O8 and O9 are however found at 0.13\,$z_T$ and 0.16\,$z_T$ from the plane of interest. Because of existing set of constraints for the design it is not possible to reduce these distances $z$ further and these two optics (O8 an O9) are therefore the most critical components of our testbed in terms of generation of amplitude-induced errors.  

\begin{table}[!ht]
\caption{Characteristics of the mirrors for the testbed with the initial and final design. For O8, the distance is estimated from the out-of-pupil plane in which the second face-sheet DM will be located (300\,mm from the first DM). The other optics are found at a distance $z$ equal to their focal length.}
\centering
\begin{tabular}{l |c c| c c| c c| c c}
\hline\hline
\textbf{Optics} & \multicolumn{2}{c|}{\textbf{Focal length}} & \multicolumn{2}{c|}{\textbf{Optics size}} & \multicolumn{2}{c|}{\textbf{Beam size}} & \multicolumn{2}{c}{\textbf{Distance $z$}}\\
 & \multicolumn{2}{c|}{\textbf{in mm}} & \multicolumn{2}{c|}{\textbf{in mm}} & \multicolumn{2}{c|}{\textbf{in mm}} & \multicolumn{2}{c}{\textbf{in $z_T$}}\\
\hline
design configuration & Initial & Final & Initial & Final & Initial & Final & Initial & Design\\
\hline
O1, O2, O5, O6 & 800 & 478 & 50.8 & 50.8 & 14.0 & 20.0 & 0.13 & 0.04\\
O3, O4 & 400 & 200 & 25.4 & 25.4 & 7.0 & 8.5 & 0.26 & 0.09\\
O7 & 400 & 210 & 25.4 & 25.4 & 7.0 & 9.0 & 0.26 & 0.08\\
O8 & 700 & 713 & 25.4 & 25.4 & 7.0 & 9.0 & 0.26 & 0.16\\
O9 & 1400 & 1597 & 50.8 & 50.8 & 14.0 & 20.0 & 0.23 & 0.13\\
O10 & 537 & 1177 & 50.8 & 50.8 & 14.0 & 20.0 & 0.09 & 0.09\\
\hline
\end{tabular}\\
\label{table:design_first_iteration}
\end{table}

\subsection{Optics surface error specifications}
\begin{figure}[!ht]
\centering
\includegraphics[scale=0.2]{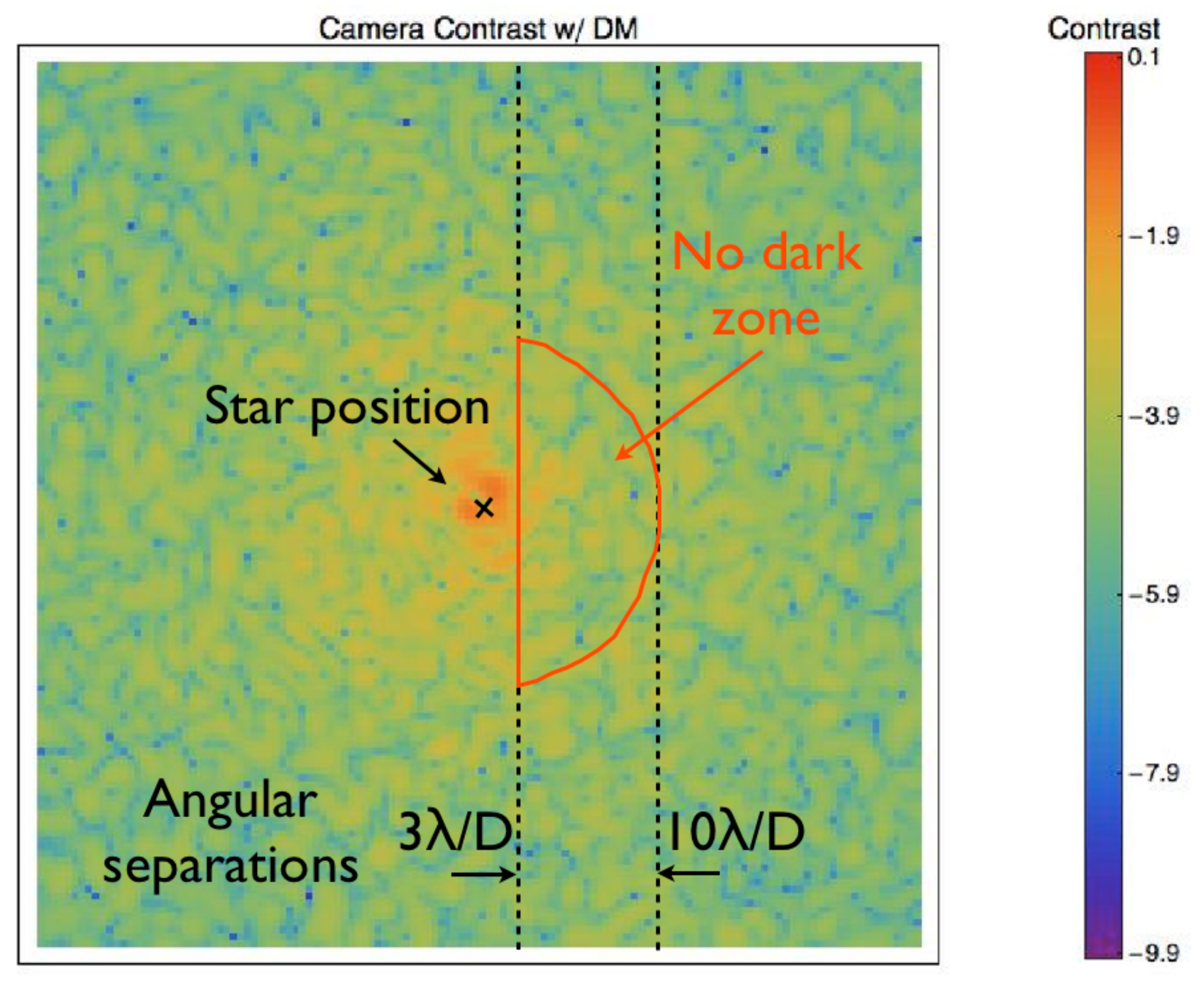}
\includegraphics[scale=0.28]{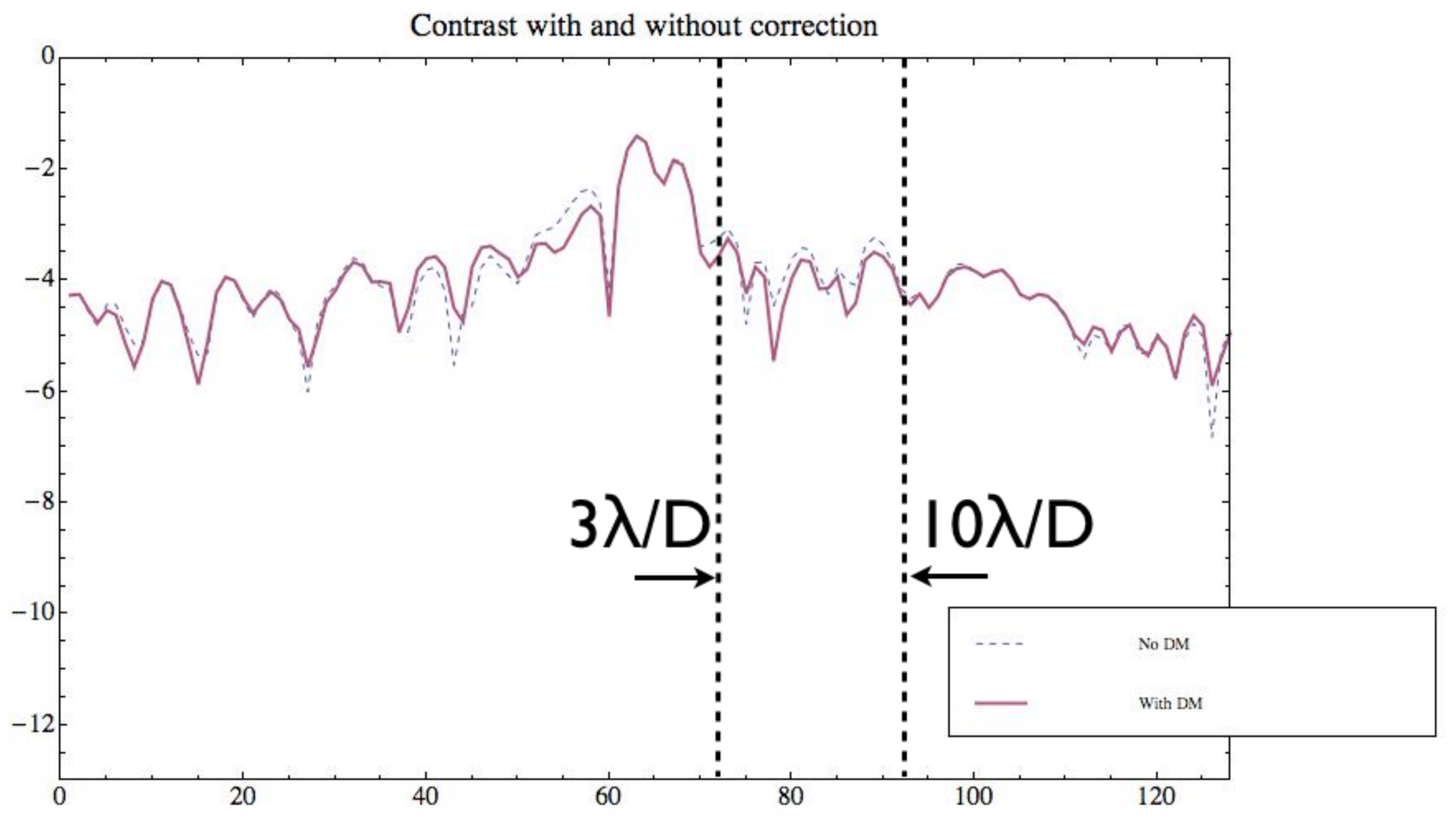}
\includegraphics[scale=0.2]{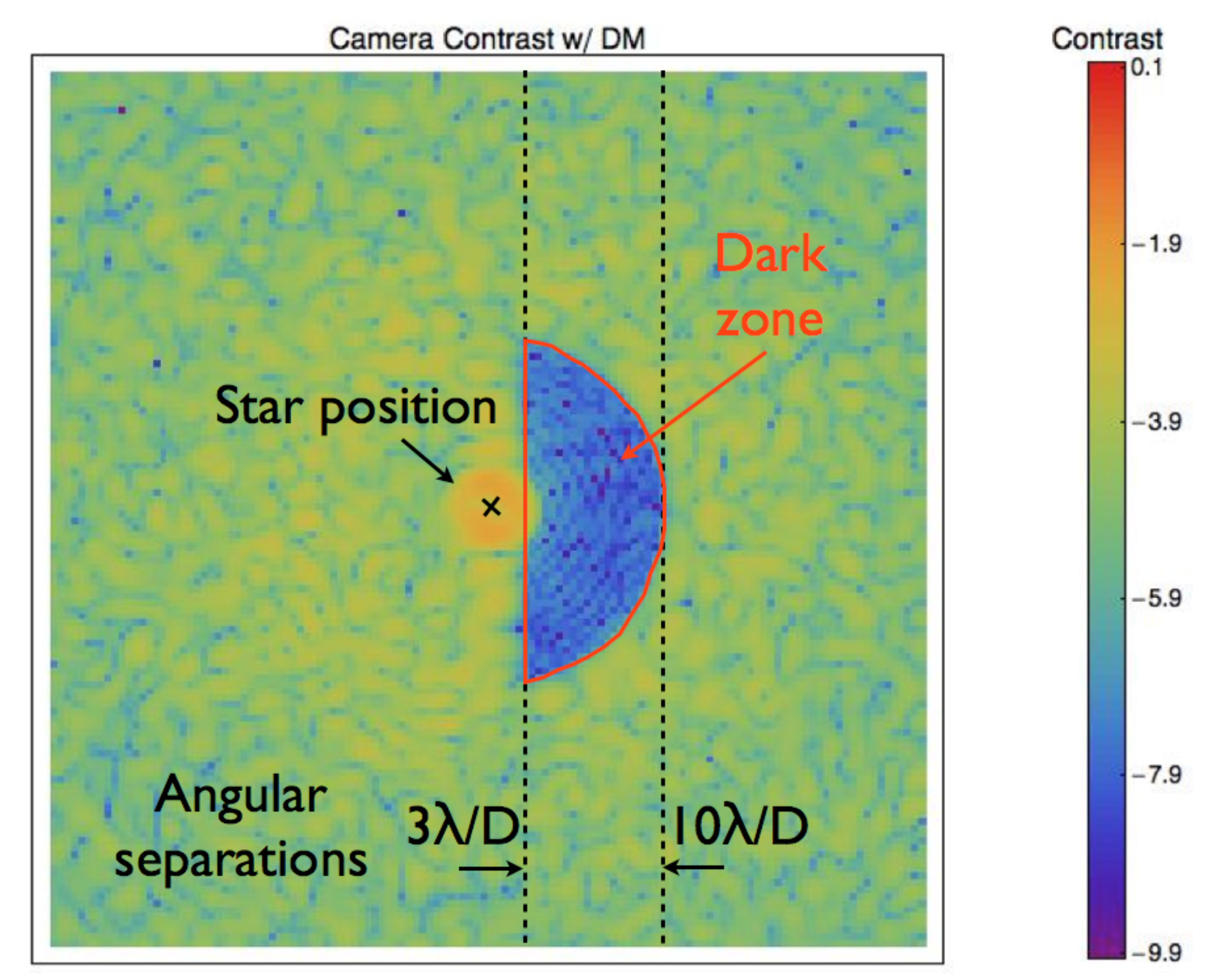}
\includegraphics[scale=0.2]{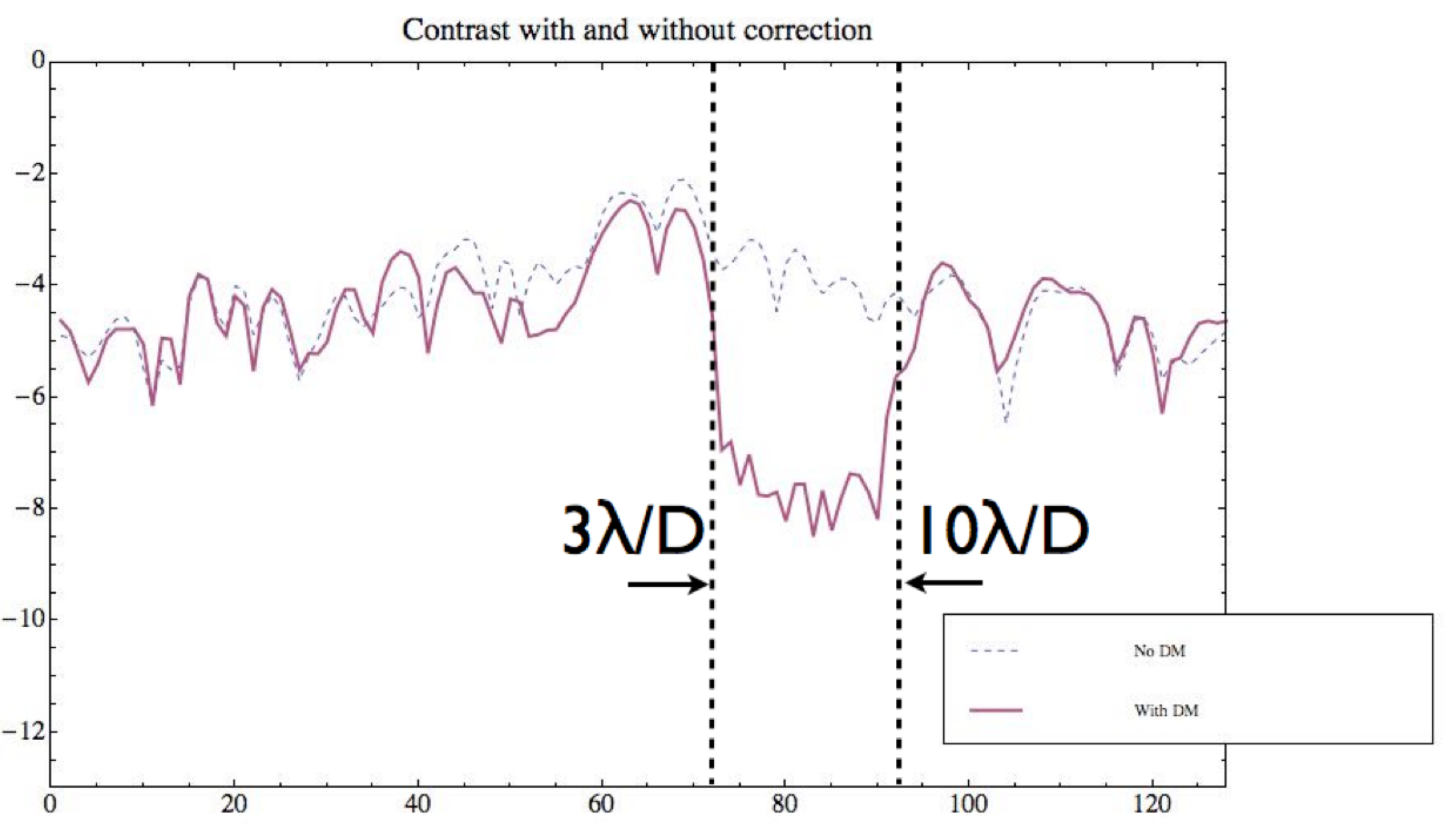}
\caption{Coronagraphic PSF obtained from end-to-end simulations with a 10\,nm rms surface errors specification for all the testbed optics, defined with the initial and final optical design configurations (top and bottom), see Table \ref{table:design_first_iteration}. \textbf{Left}: Coronagraphic image in log scale with the delimitation of the half-side zone for the wavefront shaping algorithm. A dark zone is successfully generated with the second optical design configurations (bottom). \textbf{Right}: Radial intensity profiles of the coronagraphic PSF in log scale before and after generation of the half-side dark zone.}
\label{fig:simulation_partI_PSFresult}
\end{figure}

Table \ref{table:optics_specs_first_iteration} gives the surface error specifications of the optics after the first round of simulations (i.e. without increased tolerance on low-order and without DM pre compensation) for both design configurations to reach our initial $10^{-8}$ contrast requirement in the dark zone. To understand the difference of specifications between both configurations, we repeat our simulations, assuming 10\,nm rms surface error specification for each optic, with the initial and final design configurations, and re-calculate the final coronagraphic image after application of the wavefront shaping algorithm, see Figure \ref{fig:simulation_partI_PSFresult} top and bottom. A dark hole is successfully generated with only the final testbed design, for which the optical design configuration provides large Talbot lengths.
In the initial testbed design configuration, errors are dominated by Talbot (amplitude-induced) effects. After our optimization of the optics parameters to reduce Talbot distance fractions, we switch to a different regime in which the phase effects prevail over Talbot effects. In this phase-dominated regime, the limitation on the testbed is the ability to close the loop for the stroke minimization algorithm \cite{Pueyo:09}, which puts more constraints on the quality of the initial calibration and pre-compensation. In this case, the Talbot effects are minimized and small for all of the optics.

\begin{table}[!ht]
\caption{Preliminary specifications of the optics surface errors with a $f^{-2}$ power law PSD for each optics. These results are obtained after the first round of optimizations in the end-to-end simulations (no phase pre-compensation).}
\centering
\begin{tabular}{c c c c c}
\hline\hline
design configuration & \multicolumn{3}{c}{\textbf{Surf. err. in nm rms}} & \textbf{Dark hole}\\
& \textbf{Large optics} & \textbf{Small optics} & \textbf{Optics} & \textbf{averaged}\\
& \textbf{(O2, O5, O6, O9)} & \textbf{(O3, O5, O7)} & \textbf{O8} & \textbf{intensity}\\
\hline
Initial & 10 & 2.5 & 2.5 & $5 \times 10^{-8}$\\
Final & 10 & 10 & 10 & $3 \times 10^{-8}$\\
\hline
\end{tabular}\\
\label{table:optics_specs_first_iteration}
\end{table}

Table  \ref{table:optics_specs_second_iteration} gives the final surface error specifications of the different optics for our final testbed design in the low-, mid- and high-spatial frequencies (LoF, MiF and HiF) ranges after simulations. The final results show identical error specifications for all the mirrors, following the reasoning point exposed in the previous section.

\begin{table}[!ht]
\caption{Surface error specifications for the optics of the final testbed design in different spatial frequency ranges.}
\centering
\begin{tabular}{l c c c c}
\hline\hline
\multirow{2}{*}{\textbf{Optics}} & \textbf{Focal length} & \multicolumn{3}{c}{\textbf{Surf. err. in nm rms}}\\
 & \textbf{in mm} & \textbf{LoF} & \textbf{MiF} & \textbf{HiF}\\
\hline
O2, O5, O6 & 478 & 40 & 6.4 & 5.2\\
O3, O4 & 200 & 40 & 6.4 & 5.2\\
O7 & 210 & 40 & 6.4 & 5.2\\
O8 & 713 & 40 & 6.4 & 5.2\\
O9 & 1597 & 40 & 6.4 & 5.2\\
\hline
\end{tabular}\\
\label{table:optics_specs_second_iteration}
\end{table}

\section{OPTICS REQUIREMENTS}\label{sec:requirements}
With simulation results as a starting point, we establish the final requirements for the different optics, accounting for alignment tolerance aspects.

After obtaining a specification for the optics in LoF range by simulations, we split this value into low-order spatial frequencies wavefront errors $\sigma_{WFE}$ and tolerance specification for optics alignments $\sigma_{Align}$. The total wavefront error $\sigma_{totalWFE}$ is then given by:
\begin{equation}
\sigma_{totalWFE}=\sqrt{\sigma^2_{WFE} + \sigma^2_{Align}}\,.
\end{equation}
As a starting point, we consider $\sigma_{totalWFE}$=80\,nm rms WFE, derived from the 40\,nm rms surface error specification found  by simulations. We then balance the alignment WFE and the surface error in the LoF specifications, resulting in  
\begin{equation}
\sigma_{Align}=\sigma_{Surf. err}=\sigma_{WFE}/2=35.8\,\text{nm rms}\,.
\end{equation}
Since this value is close to the standard value of $\lambda$/20 (31.7 nm) rms at He-Ne source, we adopt the latter for the LoF surface errors and the alignment wavefront errors. 
Table \ref{table:optics_specs_for_manufacturers} gives the specifications for each optic accounting for the optics alignment errors, based on our simulations. In this case the total WFE in the testbed is $\sim$150\,nm RMS without the correction using the DMs. This allows for $\lambda$/20 optics (surface error) and leaves another $\sim\lambda$/20 for alignment error budget, which translates into comfortable mechanical tolerances (typically in the range 100 to 500\,$\mu$m alignment tolerances depending on the optic and based on our more detailed alignment tolerancing study with Zemax). 
This total WFE is also at the level of our maximum acceptable requirement, in order to have sufficient Strehl ratio without correction for alignment work.
A testbed design review was conducted in May 2013. From the different discussions and iterations with potential vendors, we decided to introduce a factor two margin for manufacturing, to push back the ultimate contrast limit set by the testbed optics, to provide some margin in our error budget and also to improve the quality of PSFs with higher Strehl ratio without DM correction (useful during the alignment phase of the testbed), see Table \ref{table:optics_specs_for_manufacturers}. 

\begin{table}[!ht]
\caption{Preliminary set and final optics specifications for the manufacturers.}
\centering
\begin{tabular}{l c c |c c c | c c c}
\hline\hline
\multirow{3}{*}{\textbf{Optics}} & \textbf{Focal length} & \textbf{WFE} & \multicolumn{3}{c|}{\textbf{Preliminary specs.}} & \multicolumn{3}{c}{\textbf{Final specifications}}\\
 & \textbf{in} & \textbf{for alignment} & \multicolumn{3}{c|}{\textbf{surf. err. in nm rms}} & \multicolumn{3}{c}{\textbf{surf. err. in nm rms}}\\
 & \textbf{mm} & \textbf{in nm rms} & \textbf{LoF} & \textbf{MiF} & \textbf{HiF} & \textbf{LoF} & \textbf{MiF} & \textbf{HiF}\\
\hline
O2, O5, O6 & 478 & 31.7  & 31.7  & 6.4 & 5.2 &15.9 & 3.2 & 2.6\\
O3, O4 & 200 & 31.7 & 31.7  & 6.4 & 5.2 & 15.9 & 3.2 & 2.6\\
O7 & 210 & 31.7  & 31.7  & 6.4 & 5.2 & 15.9 & 3.2 & 2.6\\
O8 & 713 & 31.7  & 31.7  & 6.4 & 5.2 & 15.9 & 3.2 & 2.6\\
O9 & 1597 & 31.7  & 31.7  & 6.4 & 5.2 & 15.9 & 3.2 & 2.6\\
\hline
\end{tabular}\\
\label{table:optics_specs_for_manufacturers}
\end{table}


\section{OPTICAL DESIGN}\label{sec:testbed_design}
\subsection{Description}
\begin{figure}[!ht]
\centering
\includegraphics[width=\linewidth]{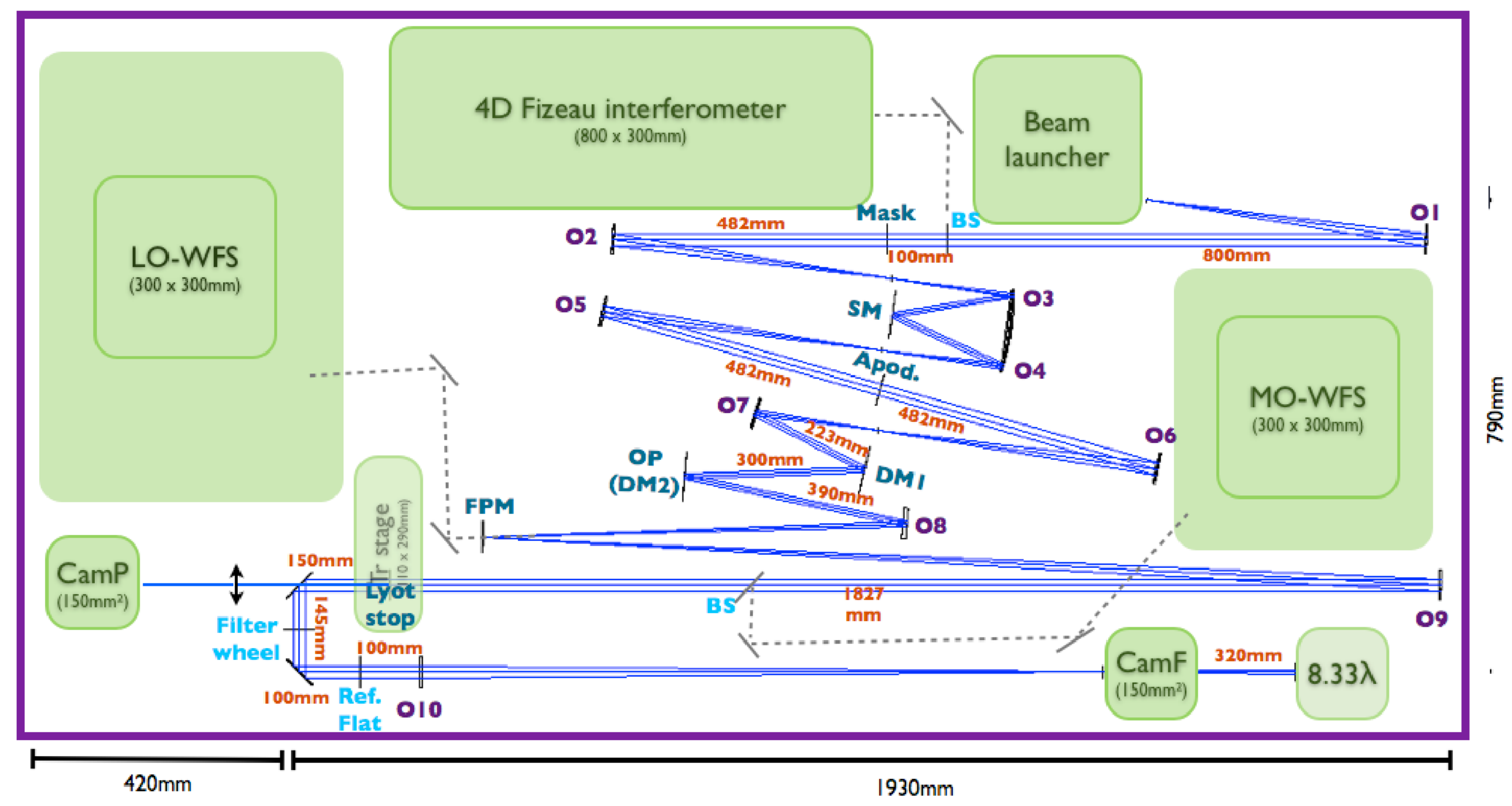}
\caption{Overview of the testbed design, composed of nine reflective optics (from O1 to O9) and one refractive optic O10 to form the image on the camera. The optics O3 and O4 are part of a one-piece parabola. The reference flat is introduced in the optical path for use with the 4D interferometer. Beamsplitters (BS) are represented in the optical path for beam injection to or extraction from the main optical path. The camF is represented at two different positions (in and out-of-focus with a 8.33$\lambda$ defocus), allowing focal plane imaging and phase retrieval applications.}
\label{fig:testbed_optical_design}
\end{figure}

Figure \ref{fig:testbed_optical_design} gives an overview of the testbed layout, which is mostly built with reflective optics for a 2.4\,m$\times$1.2\,m table space with lots of flexibility to allow future evolutions of the testbed. The design includes space for a focal plane imager with motorization for phase diversity measurements, a pupil imager after the Lyot stop, and a 4D AccuFiz interferometer for alignment and direct wavefront sensing measurements. 
The current implementation accounts for two Boston kilo-DMs, although one of them will initially be replaced by a flat mirror. We also allow room for a low-order wavefront sensor behind the FPM and a mid- or high-order wavefront sensor using a possible pick-off close to the Lyot stop or using Lyot stop reflected light \cite{2012SPIE.8450E..0NN, 2013A&A...555A..94N, 2010SPIE.7736E.179W, 2011SPIE.8126E..11W,2009ApJ...693...75G}. 
The requirement to fit on the existing optical table places significant contrasting on the testbed design. We do not use telecentric beam for the source launch because this is not a necessary requirement. Since the focal length of the first optics is small (478\,mm), using a telecentric beam would prevent us from setting the beam launcher and the 4D interferometer at the beginning of the design. The rest of the testbed however uses telecentric beams. The beam focal ratio on the focal plane mask is limited below F/90 in the presence of a 20\,mm beam size to limit the focal length for O9 and allow room behind the Lyot stop (the final f-ratio is F/80).
A refractive lens is used for O10 to image the field on the focal plane camera. This relaxes the length of the footprint of the design and leaves room for the pupil imager. Finally, two refractive beamsplitters (or removable fold mirrors) are useful before the Lyot stop: one for the injection of the 4D beam in the design, and another to pick up light for the mid- or high-order wavefront sensor.


\subsection{Optics shape}
Different shapes are explored for each optics (from O1 to O9) to find a trade-off between feasible parts at a moderate price and minimal design wavefront error ($<$ 5\,nm rms). The mirrors O1, O2, O5 and O6 are identical and easily manufacturable as off-axis parabolas (OAPs, moderately slow, small decenter). The mirrors O3 and O4 are implemented using a double bounce on a large on-axis parabola.
For the optics O7, O8, and O9, we considered different shapes (OAP, toric mirrors, spherical mirrors) and evaluated the design WFE to find a configuration that meets our requirements, see Table \ref{table:optics_shape}. Toric mirrors have been manufactured with super polishing methods and implemented with success on VLT/SPHERE \cite{2012A&A...538A.139H}. We selected the configurations 2 and 9 (O7: oap, O8: oap/toric, O9: sphere) which provide us a WFE design below 5\,nm rms. 
 
\begin{table}[!ht]
\caption{WFE design as a function of the combination of optics shape for O7, O8 and O9. A design WFE below 5\,nm rms is reached with configurations 1, 2, 5, 6, and 9. The mark $*$ underlines the selected designs.}
\centering
\begin{tabular}{c c c c c}
\hline\hline
\textbf{Config.} & \multirow{2}{*}{\textbf{O7}} & \multirow{2}{*}{\textbf{O8}} & \multirow{2}{*}{\textbf{O9}} & \textbf{WFE}\\
\#  & & & & \textbf{in nm rms}\\
\hline
1 & oap & oap & oap & 0.5\\
2* & oap & oap & sphere & 4.3\\
3 & oap & sphere & sphere & 31\\
4 & sphere & sphere & sphere & 300\\
5 & oap & oap & toric & 0.9\\
6 & oap & toric & toric & 2.7\\
7 & toric & toric & toric & 25.9\\
8 & toric & toric & sphere & 26.1\\
9* & oap & toric & sphere & 4.8\\   
\hline
\end{tabular}\\
\label{table:optics_shape}
\end{table}

\subsection{Alignment tolerancing}
We perform an analysis of alignment tolerances for the different optics of the testbed, see Table \ref{table:optics_tolerance}. The optics O3 and O4 are a double-bounce of a single on-axis parabola and therefore, the tolerance listed here is for that single optics. We find reasonable tolerancing values for the optics positioning, allowing us for some flexibility for their alignment.

\begin{table}[!ht]
\caption{Tolerances for the different optics of the testbed.}
\centering
\begin{tabular}{c c c c c c c}
\hline\hline
\multirow{3}{*}{\textbf{Optics}} & \multicolumn{6}{c}{\textbf{Tolerances $\pm$}}\\
 & \textbf{X decenter} & \textbf{Y decenter} & \textbf{Z decenter} & \textbf{tilt about X} & \textbf{tilt about Y} & \textbf{tilt about Z}\\
 & \textbf{in mm} & \textbf{in mm} & \textbf{in mm} & \textbf{in arcmin} & \textbf{in arcmin} & \textbf{in arcmin}\\ 
\hline
O1 & 1.8 & 0.4 & 0.2 & 1.3 & 10.4 & 3\\
O2 & 2.3 & 0.4 & 0.2 & 1.1 & 13.1 & 4\\
O3 / O4 & 1.3 & 0.3 & 0.1 & 2.9 & 10.4 & n.a. \\
O5 & 2.4 & 2.1 & 0.2 & 6.0 & 5.7 & 1\\
O6 & 0.8 & 0.7 & 0.2 & 3.4 & 3.3 & 4\\
O7 & 0.8 & 0.7 & 0.2 & 11.2 & 10.6 & 1\\
O8 & 2.0 & 3.9 & 0.4 & 6.0 & 5.2 & 1\\
O9 & 5.2 & 6.6 & 0.9 & 5.3 & 5.1 & 34\\
\hline
\end{tabular}\\
\label{table:optics_tolerance}
\end{table}

\section{CONCLUSIONS}\label{sec:conclusions}
In this communication, we have presented the STScI testbed and the approach adopted for its design. Our testbed aims to provide high-contrast imaging solutions that combine wavefront sensing and control, and starlight suppression strategies for complex aperture geometries. We have designed a testbed with minimal impact of the bench optics, in particular with a minimization of Talbot effects, which allows us to focus entirely on the effects of aperture geometry. Our methodology combines ray tracing optical design and end-to-end Fresnel propagation simulations to generate the specifications that place the amplitude-induced errors one order of magnitude below our final contrast requirement. A contrast limitation $\sim10^{-8}$ has been numerically determined with our iterative approach and an ultimate value better than $10^{-8}$ is expected with our final optics specifications, matching the requirement imposed at the beginning of the study. i.e. one order of magnitude below our contrast goal of $10^{-7}$ at 3\,$\lambda/D$ assuming a single Boston DM. 
The testbed has been designed with space and interfaces for future optional features to allow for innovative concepts, in particular implementations of wavefront sensors, coronagraphs or wavefront control techniques for segmented or on-axis apertures.
A design review of the testbed was performed in May 2013, allowing us to freeze our optical design and optics specifications. At the time of this communication the testbed parts are in procurement, and we expect to start the alignment operations in Fall 2013. 

\acknowledgments     
This work is supported by the National Aeronautics and Space Administration under Grant NNX12AG05G issued through the Astrophysics Research and Analysis (APRA) program (Soummer, PI). The authors acknowledge all the people involved during the process of the testbed design. In particular, they warmly thank Matt Kenworthy, Emmanuel Hugot and Marc Ferrari for fruitful discussions.

\bibliography{2013_mndiaye_spie_8864-55_v06}   
\bibliographystyle{spiebib}   

\end{document}